\documentclass[a4paper,aps,pra,reprint,superscriptaddress,floatfix]{revtex4-1}
\usepackage{amsmath} 

\usepackage{graphicx}
\usepackage[colorlinks,
	linkcolor=red,
	citecolor=blue,
	urlcolor=red]{hyperref}
\usepackage[all]{hypcap}

\usepackage{cleveref}
\usepackage{bbold}

\usepackage{xfrac}

\usepackage[xcolor={usenames,dvipsnames},draft]{changes}

\definechangesauthor[color=purple]{NRB}
\definechangesauthor[color=blue]{TJK}
\definechangesauthor[color=Green]{LDT}
\definechangesauthor[color=ProcessBlue]{AKF}
\definechangesauthor[color=Orange]{AN}
\definechangesauthor[color=brown]{AK}
\definechangesauthor[color=Mahogany]{DM}

\usepackage{mathtools}

\newcommand{\w}{\omega}
\newcommand{\W}{\Omega}

\newcommand{\meone}{\hat b_1}
\newcommand{\metwo}{\hat b_2}
\newcommand{\mejay}{\hat b_j}
\newcommand{\mwone}{\hat a_1}
\newcommand{\mwtwo}{\hat a_2}
\newcommand{\mweye}{\hat a_i}
\newcommand{\goo}{g_{11}}
\newcommand{\gto}{g_{21}}
\newcommand{\got}{g_{12}}
\newcommand{\gtt}{g_{22}}
\newcommand{\meomone}{\Omega_1}
\newcommand{\meomtwo}{\Omega_2}
\newcommand{\mwomone}{\omega_{\mathrm{c,1}}}
\newcommand{\mwomtwo}{\omega_{\mathrm{c,2}}}
\newcommand{\Gammeffone}{\Gamma_{\mathrm{eff,1}}}

\newcommand{\Gammeffjay}{\Gamma_{\mathrm{eff},j}}
\newcommand{\Gammaone}{\Gamma_{\mathrm{m,1}}}
\newcommand{\Gammatwo}{\Gamma_{\mathrm{m,2}}}
\newcommand{\Gammajay}{\Gamma_{\mathrm{m},j}}
\newcommand{\Gammaeye}{\Gamma_{\mathrm{m},i}}
\newcommand{\Gammasym}{\Gamma_{\mathrm{m}}}
\newcommand{\Gammacross}{\Gamma^{\mathrm{(cross)}}_{\mathrm{m},1}}
\newcommand{\kapone}{\kappa_1}
\newcommand{\kaptwo}{\kappa_2}
\newcommand{\kapeye}{\kappa_i}
\newcommand{\kapexone}{\kappa_{\mathrm{ex},1}}
\newcommand{\kapextwo}{\kappa_{\mathrm{ex},2}}
\newcommand{\coop}{\mathcal{C}}
\newcommand{\product}{\cdot}
\newcommand{\pump}{\mathrm{p}}

\newcommand{\mwthree}{\hat a_3}
\newcommand{\kapthree}{\kappa_3}

\newcommand{\opa}{\hat{a}}
\newcommand{\opb}{\hat{b}}
\newcommand{\oph}{\hat{H}}
\newcommand{\dg}{\dagger}

\newcommand{\eff}{\mathrm{eff}}

\newcommand{\cc}{\mathcal{C}}

\newcommand{\mrm}[1]{\mathrm{#1}}

\newcommand*\mat[1]{\begin{pmatrix}#1\end{pmatrix}}

\newcommand*\ev[1]{\langle #1 \rangle}
\newcommand*\dagg{^\dagger}

\newcommand{\id}{\mathbb{1}}

\begin{document}

\title{Nonreciprocal reconfigurable microwave optomechanical circuit}

\author{N.~R.~Bernier} 
\thanks{These authors contributed equally to this work}
\author{L.~D.~T\'{o}th}
\thanks{These authors contributed equally to this work}
\affiliation{Institute of Physics, {\'E}cole Polytechnique F{\'e}d{\'e}rale de Lausanne, 
	Lausanne 1015, Switzerland}
\author{A.~Koottandavida}
\author{M.~Ioannou}
\affiliation{Institute of Physics, {\'E}cole Polytechnique F{\'e}d{\'e}rale de Lausanne, 
	Lausanne 1015, Switzerland}
\author{D.~Malz}
\author{A.~Nunnenkamp}
\affiliation{Cavendish Laboratory, University of Cambridge, Cambridge CB3 0HE, United Kingdom}
\author{A.~K.~Feofanov}
\email{alexey.feofanov@epfl.ch}
\affiliation{Institute of Physics, {\'E}cole Polytechnique F{\'e}d{\'e}rale de Lausanne, 
	Lausanne 1015, Switzerland}
\author{T.~J.~Kippenberg}
\email{tobias.kippenberg@epfl.ch}
\affiliation{Institute of Physics, {\'E}cole Polytechnique F{\'e}d{\'e}rale de Lausanne, 
	Lausanne 1015, Switzerland}

\begin{abstract}
Devices that achieve nonreciprocal microwave transmission are ubiquitous in radar and radio-frequency communication systems, 
and commonly rely on magnetically biased ferrite materials.
Such devices are also indispensable in the readout chains of superconducting quantum circuits 
as they protect sensitive quantum systems from the noise emitted by readout electronics. 
Since ferrite-based nonreciprocal devices are bulky, lossy, and require large magnetic fields, 
there has been significant interest in magnetic-field-free on-chip alternatives,
such as those recently implemented using Josephson junctions.
Here we realise reconfigurable nonreciprocal transmission 
between two microwave modes using purely optomechanical interactions in a 
superconducting electromechanical circuit. 
We analyse the transmission as well as the noise properties of this nonreciprocal circuit.
The scheme relies on the interference in two mechanical modes that mediate coupling between microwave cavities.
Finally, we show how quantum-limited circulators can be realized with the same principle.
The technology can be built
on-chip without any external magnetic field,
and is hence
fully compatible with superconducting quantum circuits. 
All-optomechanically-mediated nonreciprocity demonstrated here can also be extended
to implement directional amplifiers,
and it forms the basis towards realising topological states of light and sound.

\end{abstract}

\date{\today}

\maketitle

Nonreciprocal devices, such as isolators, circulators, and directional amplifiers, 
exhibit altered transmission characteristics 
if the input and output channels are interchanged.
They are essential to several applications in signal processing and communication,
as they protect devices from interfering signals
~\cite{pozar_microwave_2011}.
At the heart of any such device lies an element
breaking Lorentz reciprocal symmetry for electromagnetic sources
~\cite{feynman_QED_1988,jalas_what_2013}. 
Such elements have included
ferrite materials~\cite{auld_synthesis_1959,milano_y-junction_1960,fay_operation_1965}, 
magneto-optical materials~\cite{aplet_faraday_1964,shirasaki_compact_1982,sato_lens-free_1999,bi_on-chip_2011}, 
optical nonlinearities~\cite{manipatruni_optical_2009,fan_all-silicon_2012,guo_on-chip_2016},
temporal modulation~\cite{anderson_reciprocity_1965,yu_complete_2009,lira_electrically_2012,kang_reconfigurable_2011,estep_magnetic-free_2014,peng_parity-time-symmetric_2014},
chiral atomic states~\cite{scheucher_quantum_2016}, 
and physical rotation
~\cite{fleury_sound_2014}.
Typically, a commercial nonreciprocal microwave apparatus exploits 
ferrite materials and magnetic fields, 
which leads to a propagation-direction-dependent phase shift for different field polarizations. 
A significant drawback of such devices is that they 
are ill-suited for sensitive superconducting circuits,
since their strong magnetic fields are disruptive and require heavy shielding.
In recent years, the major advances in quantum superconducting circuits
~\cite{Devoret_2013},
that require isolation from noise emanating from readout electronics,
have led to a significant interest in nonreciprocal devices operating at the microwave frequencies 
that dispense with magnetic fields and can be integrated on-chip.

\begin{figure}
  \includegraphics[width=0.48\textwidth]
  {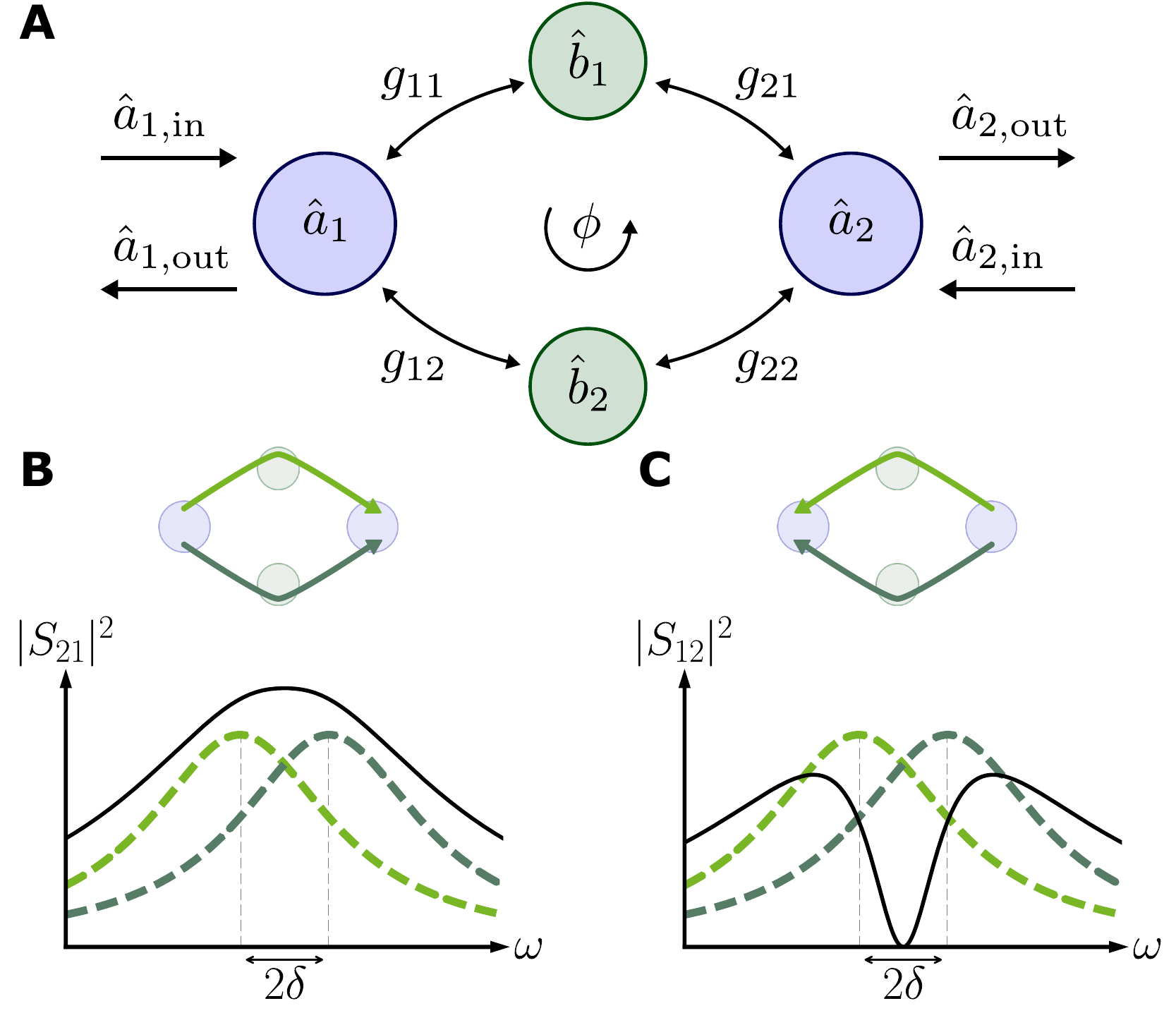}
  \caption{
  \textbf{Optomechanical nonreciprocal transmission via interference of two asymmetric dissipative coupling pathways.}
  \textbf{A}.~Two microwave modes
  $\mwone$ and $\mwtwo$ 
  are coupled via two mechanical modes $\meone$ and $\metwo$
  through optomechanical
  frequency conversion (as given by the coupling constants $\goo,\gto,\got,\gtt$).
 Nonreciprocity is based on the interference between the two optomechanical (conversion) pathways 
  $\goo,\gto$ and $\got,\gtt$, in the presence of a suitably chosen phase difference $\phi$ between the coupling constants as well as the deliberate introduction of an asymmetry in the pathways.
  \textbf{B-C.}~
  The symmetry between the pathways can be broken
  by off-setting the optomechanical transmission windows 
  through each mechanical mode
  (dashed lines in dark and light green)
  by a frequency difference $2\delta$.
  Each single pathway, in the absence of the other mode,
  is described by \cref{eq:SingleConv}.
  In the forward direction 
  (\textbf{B}),
  the two paths
  interfere constructively, allowing transmission 
  and a finite scattering matrix element $S_{21}$ on resonance with the first microwave cavity.
  In contrast, in the backward direction 
  (\textbf{C}),
  the paths interfere destructively,
  such that $S_{12}\approx0$, thereby isolating port 1 from port 2 on resonance with the second microwave cavity.
  The isolation bandwidth is determined by the intrinsic dissipation rate of the mechanical modes.
  \label{fig:figure1}
   }
\end{figure} 

As an alternative to ferrite-based nonreciprocal technologies,
several approaches have been pursued
towards nonreciprocal microwave chip-scale devices.
Firstly, the modulation in time of the
parametric couplings between modes of a network
can simulate
rotation about an axis, creating an artificial  magnetic field
~\cite{anderson_reciprocity_1965,estep_magnetic-free_2014,kerckhoff2015,ranzani_geometric_2014} 
rendering the system nonreciprocal with respect to the ports. 
Secondly, phase matching of a parametric interaction 
can lead to  nonreciprocity, 
since the signal only interacts with the pump when copropagating with it
and not in the opposite direction.
This causes  travelling-wave amplification to be directional
~\cite{ranzani_geometric_2014,white2015,macklin_nearquantum-limited_2015,hua2016}.
Phase-matching-induced nonreciprocity can also occur 
in optomechanical systems~\cite{cavity_optomechanics_RMP,Bowen_Milburn_Quantum_2015}, where parity considerations 
for the interacting spatial modes apply
~\cite{hafezi_optomechanically_2012,shen_experimental_2016,ruesink2016}.
Finally, interference in parametrically coupled multi-mode systems can be used. 
In these systems nonreciprocity arises due to interference 
between multiple coupling pathways 
along with dissipation in ancillary modes~\cite{ranzani_graph-based_2015}. 
Here dissipation is a key resource to break reciprocity,
as it forms a flow of energy always leaving the system,
even as input and output are interchanged. 
It has therefore been viewed as reservoir engineering
~\cite{metelmann_nonreciprocal_2015}. 
Following this approach, nonreciprocity has recently been demonstrated 
in Josephson-junctions-based microwave circuits
~\cite{sliwa_reconfigurable_2015,lecocq_nonreciprocal_2017} 
and in a photonic-crystal-based optomechanical circuit
~\cite{fang_generalized_2017}.
These realisations and theoretical proposals to achieve nonreciprocity in multi-mode systems rely on a \emph{direct, coherent} coupling between 
the \emph{electromagnetic} input and output modes.

Here, in contrast, we describe a scheme to attain reconfigurable nonreciprocal transmission
without a need for any direct coherent coupling between input and output modes,
using purely optomechanical interactions~\cite{cavity_optomechanics_RMP,Bowen_Milburn_Quantum_2015}.  
This scheme neither requires cavity-cavity interactions nor phonon-phonon coupling, 
which are  necessary for the recently demonstrated 
optomechanical nonreciprocity in the optical domain
~\cite{fang_generalized_2017}.
Two paths of transmission between the microwave modes are established,
through two distinct mechanical modes.
Interference between those paths with differing phases
forms the basis of the nonreciprocal process.
In fact, due to the finite quality factor 
of the intermediary mechanical modes,
both conversion paths between the electromagnetic modes are partly dissipative in nature.
Nonreciprocity is in this case only possible by breaking the symmetry
between the two dissipative coupling pathways.
We describe the mechanism in detail below,
shedding some light on the essential ingredients for nonreciprocity using this approach.

\begin{figure*}[t]
  \includegraphics[width=0.78\textwidth]
  {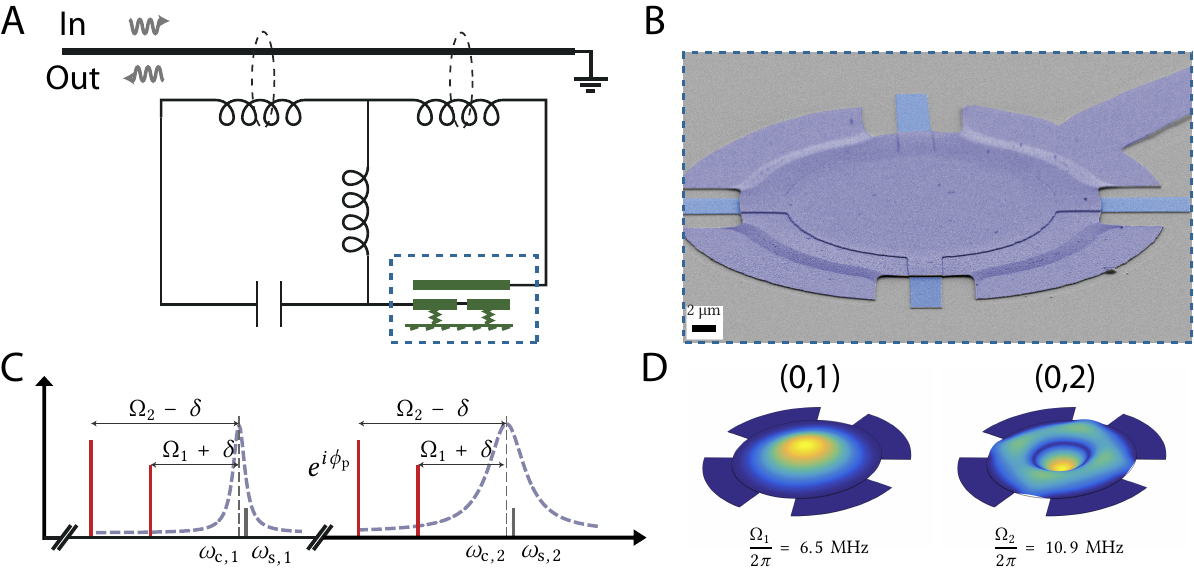}
  \caption{
  \label{fig:figure2} 
  \textbf{Implementation of a superconducting microwave circuit optomechanical device for nonreciprocity.}
  \textbf{A}.~A superconducting circuit featuring two electromagnetic modes in the microwave domain is capacitively coupled to a mechanical element
  and inductively coupled to a microstrip feedline. 
  The end of the feedline is grounded and the circuit is measured in reflection.
  \textbf{B}.~Scanning electron micrograph of the drum-head-type vacuum gap capacitor (gap distance below $50$ nm), made from aluminium on a sapphire substrate.  
   \textbf{C}.~Frequency domain schematic of the microwave pump setup to achieve nonreciprocal mode conversion.
   Microwave pumps (red bars) are placed at the lower motional sidebands - corresponding to the two mechanical modes - of both microwave resonances (dashed purple lines). 
   The pumps are detuned from the exact sideband condition by $\pm \delta = 2\pi \product 18$~kHz, creating two optomechanically induced transparency windows detuned by $2\delta$ from the microwave
   resonance frequencies. The phase $\phi_\mathrm{p}$ of one the pumps is tuned. The propagation of an incoming signal (grey bar) in the forward and backward direction depends on this phase and nonreciprocal microwave transmission can be achieved.
   \textbf{D}.~ Finite-element simulation of the fundamental $(0,1)$ and second order radially symmetric $(0,2)$ mechanical modes, which are exploited as intermediary dissipative modes to achieve nonreciprocal microwave conversion.
   }
\end{figure*} 

\begin{figure*}[t]
  \includegraphics[width=0.98\textwidth]
  {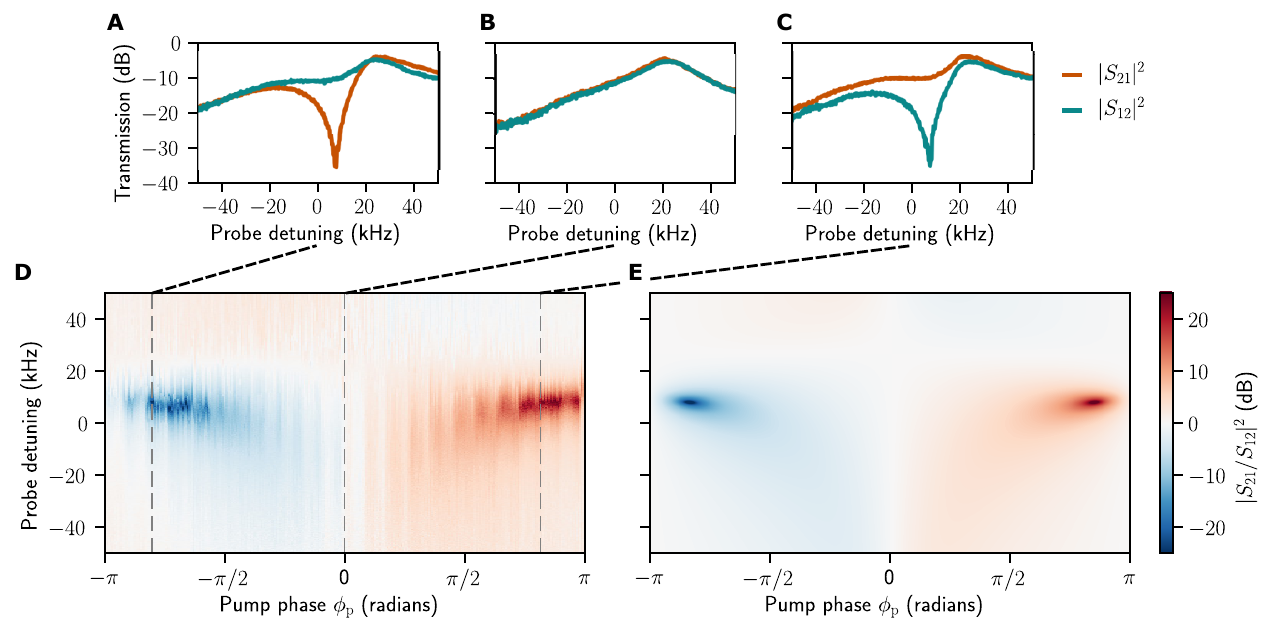}
  \caption{
  \textbf{Experimental demonstration of nonreciprocity.}
  \textbf{A}-\textbf{C}.~Power transmission
   between modes 1 and 2 as a function of probe detuning,
  shown in both directions for pump phases 
  $\phi_\mathrm{p} = -0.8\pi, 0, 0.8\pi$ radians.
  Isolation of more than 20 dB in 
  the forward (\textbf{C}) and backward (\textbf{A}) directions 
  is demonstrated, 
  as well as reciprocal behaviour (\textbf{B}).
  \textbf{D}.~The ratio of transmission 
  $|S_{21}/S_{12}|^2$, representing a measure of nonreciprocity,
  is shown as a function of pump phase $\phi_\mathrm{p}$ and probe detuning.
  Two regions of nonreciprocity develop, with isolation in each direction.
  The system is reconfigurable as the direction of isolation
  can be swapped by taking 
  $\phi_\pump \rightarrow - \phi_\pump$.
  \textbf{E}.~Theoretical ratio of transmission
  from \cref{eq:nonrep}, calculated
  with independently estimated experimental parameters.
  The theoretical model includes effectively lowered cooperativities
  for the mechanical mode $\meone$
  due to cross-damping (optomechanical damping of the lower frequency mechanical mode
  by the pump on the sideband of the higher frequency mechanical mode) acting as an extra loss channel.
  \label{fig:figure3}
   }
\end{figure*} 
\begin{figure*}[t]
  \includegraphics
  {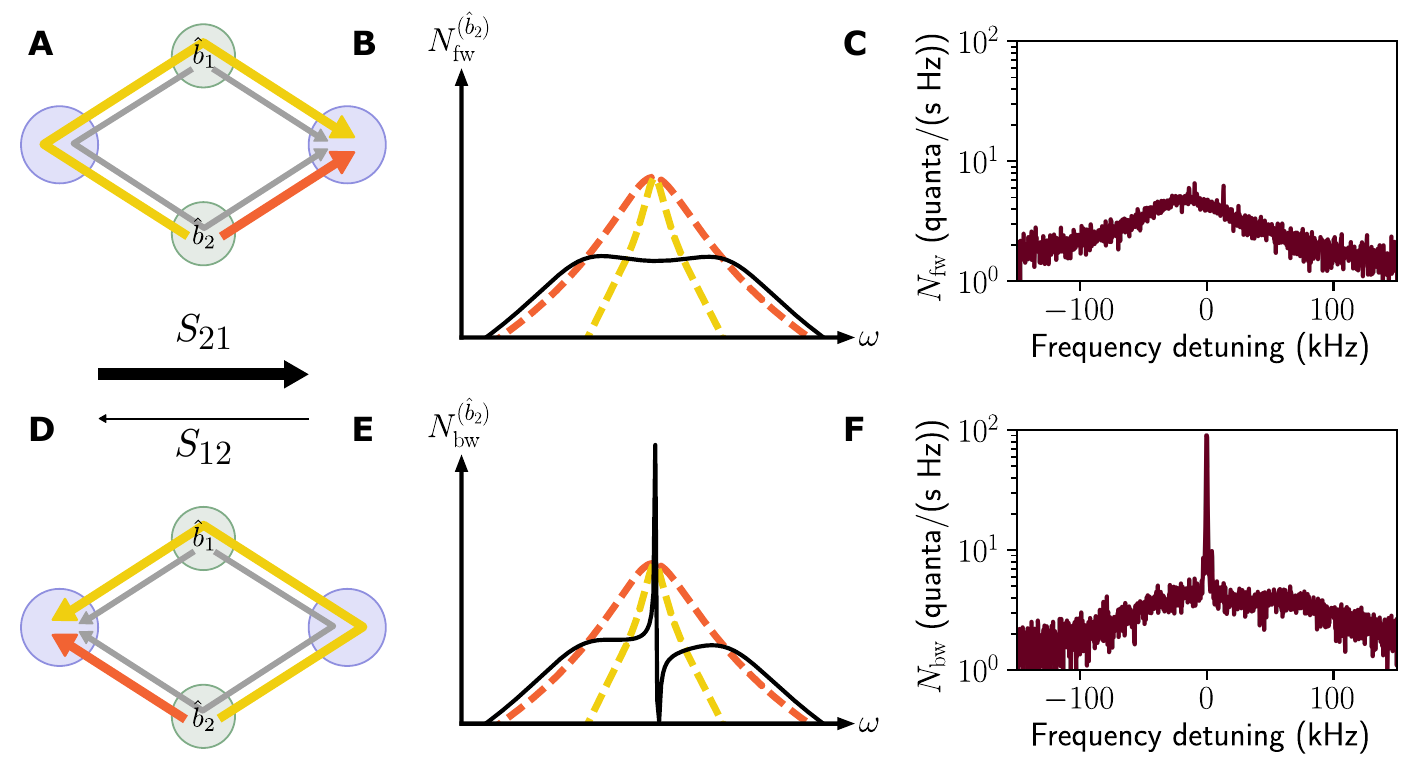}
  \caption{
  \textbf{Asymmetric noise emission of the nonreciprocal circuit.}
  The noise emission is mainly due to mechanical thermal noise,
  that is converted through two paths to the microwave modes.
  The resulting interference creates a different noise pattern
  in the forward (\textbf{A-C})
  and the backward (\textbf{D-F})
  directions when the circuit is tuned as an isolator from mode
  $\mwone$ to $\mwtwo$.
  \textbf{A, D}.~The two possible paths for the noise are shown for each
  mechanical mode.
  For $\metwo$, the direct path (orange)
  and the indirect path going through mode $\meone$ (yellow)
  are highlighted.
  \textbf{B, E}.~Each path on its own would result in a wide
  noise spectrum that is equally divided between the two microwave cavities
  (dashed yellow and orange lines).
  When both paths are available, however, the noise interferes differently
  in each direction.
  In the backward direction (\textbf{E}),
  a sharp interference peak appears, of much larger amplitude
  than the broad base.
  The theoretical curves (on an arbitrary logarithmic scale)
  are shown for the symmetric case ($\Gammaone=\Gammatwo$) 
  and for the single mode $\metwo$.
  Note that for the mode $\meone$, the shape of the asymmetric peak
  in the backward noise would be the mirror image.
  \textbf{C, F}.~
  Measured output spectra of modes 
  $\mwtwo$ (\textbf{C}) and
  $\mwone$ (\textbf{F}),
  calibrated to show the photon flux leaving the circuit.
  Because cross-damping provides extra cooling for the mode $\meone$,
  the thermal noise of $\metwo$ is expected to dominate.
  \label{fig:figure4}
   }
\end{figure*} 

\begin{figure}[t]
  \includegraphics
  {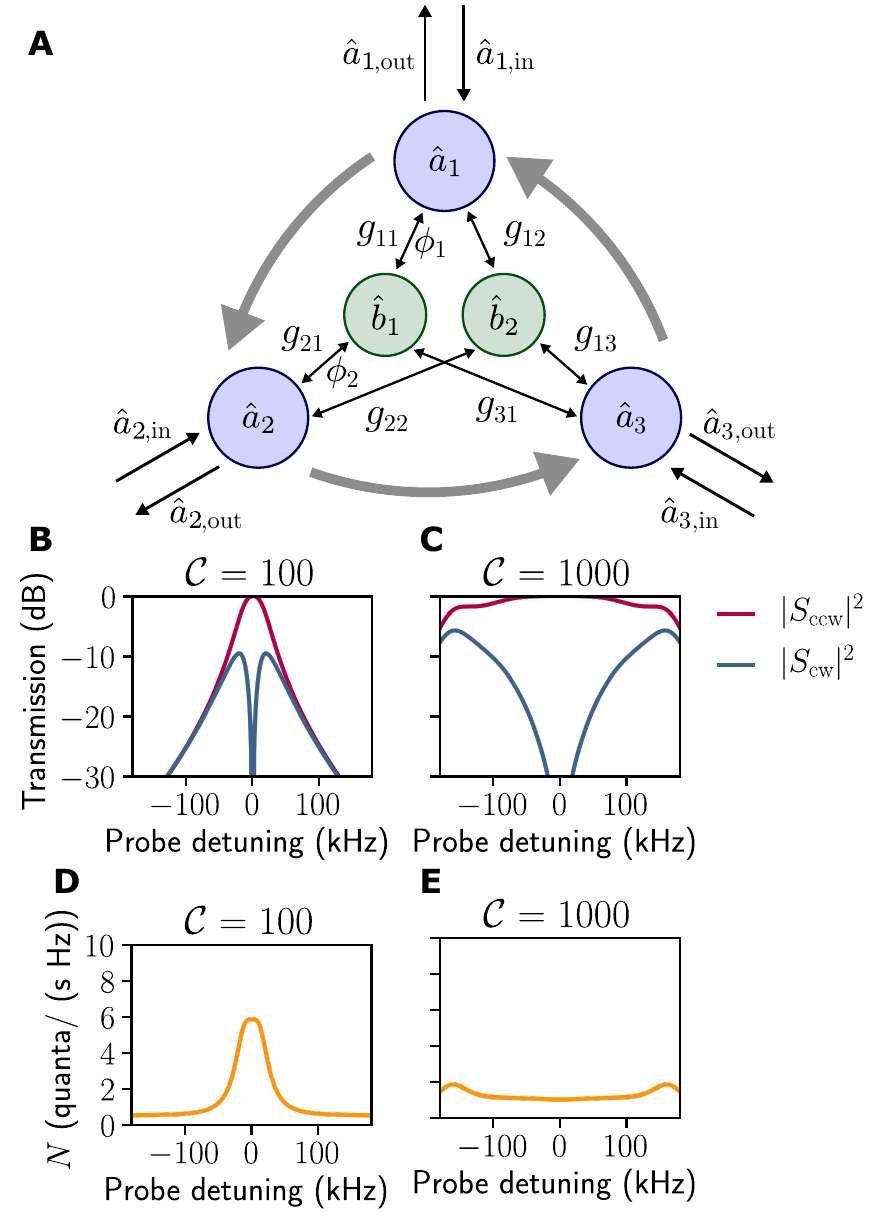}
  \caption{
  \textbf{Proposal for a microwave optomechanical circulator.}
  \textbf{A}.~
  With a third microwave mode $\mwthree$ coupled
  to the same two mechanical oscillators,
  circulation can be achieved between the three microwave cavities.
  The circuit now involves two independent loops, with two phases 
  $\phi_1$ and $\phi_2$
  that can be tuned with the phases associated with
  $\gto$ and $\goo$, respectively.
  \textbf{B-C}.~
  The theoretical transmission in the circulating direction (counter-clockwise)
  and the opposite direction (clockwise)
  are shown for two values of the cooperativity $\coop$.
  The isolation bandwidth scales with  $\coop$ and
  is only limited by the energy decay rates of the microwave modes.
  Experimentally realistic parameters are chosen
  with overcoupled cavities of energy decay rates
  $\kapone = \kaptwo = \kapthree = 2\pi \product 200$~kHz
  and $\Gammaone=\Gammatwo = 2\pi \product 100$~Hz.
  \textbf{D-E}.~%
  Noise emission spectra for the same two cooperativities,
  for $\bar n_{\mathrm{m},1} = \bar n_{\mathrm{m},2} = 800$.
  Note that for the circulator the noise is symmetric for all
  the cavities, and that it decreases with increasing cooperativity.
  \label{fig:figure5}
   }
\end{figure} 
We first theoretically model our system to reveal how nonreciprocity arises.
We consider two microwave modes (described by their annihilation operators
$\mwone$, $\mwtwo$)
having resonance frequencies $\mwomone$, $\mwomtwo$ 
and dissipation rates $\kapone$, $\kaptwo$,
which are coupled to two mechanical modes
(described by the annihilation operators $\meone$, $\metwo$) having resonance frequencies $\meomone$, $\meomtwo$ 
and dissipation rates $\Gammaone$, $\Gammatwo$
(\cref{fig:figure1}A).
The radiation-pressure-type
optomechanical interaction 
has the form~\cite{cavity_optomechanics_RMP,Bowen_Milburn_Quantum_2015}
$g_{0,ij} \mweye^\dagger \mweye (\mejay + \mejay^\dagger)$
(in units where $\hbar=1$),
where $g_{0,ij}$ designates
the vacuum optomechanical coupling strength
of the $i^{\mathrm{th}}$  microwave mode  to the $j^{\mathrm{th}}$ mechanical mode. 
Four microwave tones are applied, close to  each of the 
two lower sidebands of the two microwave modes, with detunings of
$\Delta_{11}=\Delta_{21}=-\meomone - \delta$
and
$\Delta_{12}=\Delta_{22} =-\meomtwo + \delta$
(\cref{fig:figure2}C).
We linearise the Hamiltonian, neglect counter-rotating terms, and write it in a rotating frame with respect to the mode frequencies
\begin{multline}
  H = 
  -\delta \, \meone^\dagger \meone
  +
  \delta \, \metwo^\dagger \metwo
  +
  \goo  ( \mwone \meone^\dagger + \mwone^\dagger \meone )
  +
  \gto  ( \mwtwo \meone^\dagger + \mwtwo^\dagger \meone )
  \\
  +
  \got  ( \mwone \metwo^\dagger + \mwone^\dagger \metwo )
  +
  \gtt  (  e^{i\phi}\mwtwo  \metwo^\dagger + e^{-i\phi}\mwtwo^\dagger \metwo )
  \label{eq:Ham}
\end{multline}
where $\mweye$ and $\mejay$ are redefined to be
the quantum fluctuations around the linearised mean fields.
Here
$g_{ij} = g_{0,ij} \sqrt{n_{ij}}$ are the field-enhanced optomechanical coupling strengths,
where  $n_{ij}$ is the contribution to the mean intracavity photon number 
due to the drive with detuning $\Delta_{ij}$.
Although in principle each coupling is complex, without loss of generality we can take all to be real except the one between 
$\mwtwo$ and $\metwo$
with a complex phase $\phi$. 

We start by considering frequency conversion 
through a single mechanical mode.  
Neglecting the noise terms,
the field exiting the cavity $\mwtwo$
is given by
$\mwtwo{}_{,\mathrm{out}} 
= 
S_{21} \mwone{}_{,\mathrm{in}}
+
S_{22} \mwtwo{}_{,\mathrm{in}}
$,
which defines the scattering matrix
$S_{ij}$.
For a single mechanical pathway, setting
$\got =\gtt =0$ and $\delta=0$,
the scattering matrix between input and output mode becomes
\begin{equation}
  S_{21}(\omega)
  =
  \sqrt{
  \frac{\kapexone \kapextwo}{\kapone \kaptwo}}
  \frac{\sqrt{ \coop_{11} \coop_{21} } \Gammaone}{
  \frac{\Gammeffone}{2}
  - i \omega
  },
  \label{eq:SingleConv}
\end{equation}
where 
$\kapexone$, $\kapextwo$ denote the external coupling rates 
of the microwave modes to the feedline,
and the (multiphoton) cooperativity for each mode pair is defined as
$\coop_{ij}
=
4 g_{ij}^2 / (\kappa_i \Gammajay)
$.
Conversion occurs within the modified mechanical response over 
an increased bandwidth 
$\Gammeffone = 
\Gammaone   
\left( 1 + \coop_{11} + \coop_{21} \right)$.
This scenario, where a mechanical oscillator mediates frequency conversion between electromagnetic modes, 
has recently been demonstrated~\cite{lecocq_mechanically_2016} with a microwave optomechanical  circuits~\cite{teufel_circuit_2011}%
, and moreover used to create a bidirectional link between a microwave and an optical mode~\cite{andrews_bidirectional_2014}.
Optimal conversion, limited by internal losses in the microwave cavities,
reaches at resonance
$|S_{21}|^2_\mathrm{max} = \frac{\kapexone \kapextwo}{\kapone \kaptwo}$
in the limit of large cooperativities
$\coop_{11} = \coop_{21} \gg 1$.

We next describe nonreciprocal transmission of 
the full system with both mechanical modes.
We consider the ratio of transmission amplitudes
given by
 \begin{equation}
  \frac{S_{12} (\omega)}{S_{21}(\omega)} 
  = 
  \frac{
	\goo 
    \chi_1 (\omega)
    \gto 
	+
	\got 
    \chi_2 (\omega)
    \gtt 
	e^{+i\phi}
	}{
	\goo 
    \chi_1 (\omega)
    \gto 
	+
	\got 
    \chi_2 (\omega)
    \gtt 
	e^{-i\phi}
	}
 \label{eq:nonrep}
\end{equation}
with the mechanical susceptibilities defined as
$\chi_1^{-1} (\omega) = \Gammaone/2 - i \left( \delta + \omega \right)$
and
$\chi_2^{-1} (\omega) = \Gammatwo/2 + i \left( \delta - \omega \right)$.
Conversion is nonreciprocal
if the above expression has an magnitude that 
differs from 1.
If $S_{21}$ and $S_{12}$ differ only by a  phase,
it can be eliminated
by a redefinition of either $\mwone$ or $\mwtwo$~\cite{ranzani_geometric_2014,ranzani_graph-based_2015}.
Upon a change in conversion direction,
the phase $\phi$ of the \emph{coherent} coupling (between the microwave  and mechanical mode) is conjugated,
while the complex phase associated with the response of the \emph{dissipative} mechanical modes remains unchanged.
Physically, scattering from $1\to2$ is related to scattering from $2\to1$ via time-reversal, which conjugates phases due to coherent evolution of the system. Dissipation is untouched by such an operation and thus remains invariant.
Indeed, the mechanical dissipation is an essential ingredient  
for the nonreciprocity to arise in this system, but not sufficient on its own. 
In fact, if we align the frequency conversion windows corresponding 
to the two mechanical modes by setting $\delta=0$,
the system becomes reciprocal 
on resonance ($\omega=0$),
since there is no longer any phase difference between 
numerator and denominator.
This situation corresponds to two symmetric pathways resulting 
from purely dissipative couplings; they can interfere only in a reciprocal way.

We study the conditions for isolation,
when backward transmission $S_{12}$ vanishes
while forward transmission $S_{21}$ is non-zero.
A finite offset $2\delta$ between the 
mechanical conversion windows
causes an intrinsic phase shift for a signal
on resonance ($\omega=0$)
travelling one path compared to the other,
as it falls either on the red or the blue side of each mechanical resonance.
The coupling phase $\phi$ is then adjusted to cancel propagation
in the backward direction $S_{12}$ 
(\cref{fig:figure1}C),
by cancelling the two terms in the numerator of \cref{eq:nonrep}.
In general, there is always a frequency $\omega$
for which 
$
  |\goo \chi_1 (\omega) \gto |
  =
  |\got \chi_2 (\omega) \gtt |
$,
such that the phase  $\phi$ can be tuned to 
cancel transmission in one direction.
Specifically, for  two mechanical modes with identical decay rates 
$(\Gammaone = \Gammatwo = \Gammasym)$ and symmetric couplings
($\goo\gto = \got\gtt$),
we find that 
transmission from ports 2 to 1 vanishes on resonance if 
\begin{equation}
  \frac{\Gammasym}{2\delta} 
  = 
  \tan
  \frac{\phi}{2}
  .
  \label{eq:PhiCond}
\end{equation}
The corresponding terms of the denominator
will have a different relative phase, 
and the signal will add  constructively instead, in the forward direction
(\cref{fig:figure1}B).
The device thus acts as an isolator from $\mwone$ to $\mwtwo$, realised without relying on 
Josephson-junctions
 ~\cite{sliwa_reconfigurable_2015,lecocq_nonreciprocal_2017}.
 We now describe the conditions to minimise insertion loss
of the isolator in the forward direction.
Still considering the symmetric case,
the cooperativity is set to be the same for all modes
($\coop_{ij} = \coop$).
For a given separation $\delta$,
transmission on resonance ($\omega=0$) in the isolating direction 
has the maximum
\begin{equation}
  \left| S_{21}\right|^{2}_{\mathrm{max}}
  =
  \frac{\kapexone \kapextwo}{\kapone \kaptwo}
  \left( 
  1
  -
  \frac{1}{2\coop}
  \right)
  \label{eq:maxtransmission}
\end{equation}
for a cooperativity
$\coop = 1/2 + 2 \delta^2 / \Gammasym^2$.
As in the case for a single mechanical pathway
in \cref{eq:SingleConv},
for large cooperativity 
the isolator can
reach an insertion loss
only limited by the internal losses of the microwave cavities.

The unusual and essential role of dissipation in this nonreciprocal scheme 
is also apparent in the analysis of the bandwidth of the isolation. 
Although the frequency conversion through a single mechanical mode
has a bandwidth $\Gammeffjay$ (c.f.\ \cref{eq:SingleConv}), 
caused by the optomechanical damping of the pumps on the lower sidebands,
the nonreciprocal bandwidth is set by 
the \emph{intrinsic} mechanical damping rates.
Examination of \cref{eq:nonrep} reveals that
nonreciprocity originates from the interference
of two mechanical susceptibilities of 
widths $\Gammajay$. 
One can conclude that the intrinsic mechanical dissipation, 
which takes energy out of the system regardless of the transmission direction,
is an essential ingredient for the nonreciprocal behaviour reported here, as discussed previously ~\cite{ranzani_graph-based_2015,metelmann_nonreciprocal_2015}.
In contrast, optomechanical damping works symmetrically between input and output modes.
By increasing the coupling rates, using higher pump powers,
the overall conversion bandwidth increases, while
the nonreciprocal bandwidth stays unchanged.

We experimentally realise this nonreciprocal scheme
using a superconducting circuit optomechanical system in 
which mechanical motion is capacitively coupled to a multi-mode microwave circuit~\cite{teufel_circuit_2011}.
The circuit, schematically shown in \cref{fig:figure2}A,
supports two electromagnetic modes with resonance frequencies
$(\mwomone,\mwomtwo)=2\pi \product (4.1, 5.2)$ GHz and energy decay rates $(\kapone, \kaptwo) = 2\pi \product (0.2, 3.4)$ MHz,
both of them coupled to the same vacuum-gap capacitor.
We utilise the fundamental and second order radially symmetric $(0,2)$ modes of the capacitor's
mechanically compliant top plate~\cite{cicak_low-loss_2010} 
(cf.\ \cref{fig:figure2}B and D)
with resonance frequencies $(\meomone, \meomtwo)=2\pi \product (6.5, 10.9)$ MHz,
intrinsic energy decay rates $(\Gammaone, \Gammatwo)=2\pi \product (30, 10)$ Hz
and optomechanical vacuum coupling strengths
$(g_{0,11}, g_{0,12})=2\pi \product (91,12)$ Hz, respectively 
(with $g_{0,11}\approx g_{0,21}$ and $g_{0,12}\approx g_{0,22}$, i.e.~the two microwave cavities are symmetrically coupled to the mechanical modes).
The device is placed at the mixing chamber of a dilution refrigerator at 200 mK and all four incoming pump tones
are heavily filtered and attenuated to eliminate Johnson and phase noise 
(details are published elsewhere~\cite{toth_dissipative_2016}).
We establish a parametric coupling between the two electromagnetic and the two mechanical modes by introducing four microwave pumps
with frequencies slightly detuned from the lower motional sidebands of the resonances, as shown in \cref{fig:figure2}C and as discussed above.
An injected probe signal $\omega_{\rm{s1(s2)}}$ around the lower (higher) frequency microwave mode is then measured in reflection using a vector network analyser.

Frequency conversion in both directions,
$|S_{21}(\omega)|^2$ and $|S_{12}(\omega)|^2$,
are measured and compared in 
\cref{fig:figure3}\mbox{A-C}.
The powers of the four pumps are chosen such that the associated 
individual cooperativities 
are given by
$\mathcal{C}_{11}=520$, $\mathcal{C}_{21} = 450$, $\mathcal{C}_{12}=1350$ and $\mathcal{C}_{22}= 1280$.
The detuning from the lower motional sidebands is set to $\delta = 2\pi \product 18$~kHz.
By pumping both cavities on the lower sideband associated with the same mechanical mode,
a signal injected on resonance with one of the modes will be frequency converted to the other mode.
This process can add negligible noise, when operating with sufficiently high cooperativity, 
as demonstrated recently~\cite{lecocq_mechanically_2016}.
In the experiment, the four drive tones are all phase-locked and 
the phase of one tone $\phi_\mathrm{p}$ is varied continuously from 
$-\pi$ to $\pi$.
The pump phase is linked to the coupling phase $\phi$
by a constant offset, in our case $\phi_\pump \approx \phi + \pi$.
Between the two transmission peaks corresponding to each 
mechanical mode,
a region of nonreciprocity develops,
depending on the relative phase $\phi_\mathrm{p}$. 

The amount of reciprocity that occurs in this process is quantified and measured by the 
ratio of forward to backward conversion $|S_{21}/S_{12}|^2$. 
\Cref{fig:figure3}D shows this quantity as a function
of probe detuning and the relative pump phase.
Isolation of more than 20 dB is demonstrated in each direction
in a reconfigurable manner, i.e.\ 
the direction of isolation can be switched
by taking $\phi_\pump \rightarrow - \phi_\pump$,
as expected from \cref{eq:PhiCond}.
The ideal theoretical model, which takes into account $\Gammaone \neq \Gammatwo$,
predicts that
the bandwidth of the region of nonreciprocity is commensurate with 
the arithmetic average of the bare mechanical dissipation rates, 
$\sim 2\pi \product  20$~Hz.
However, given the significantly larger coupling strength
of the fundamental mechanical mode compared to
the second order mode, and
that $\kaptwo / \Omega_{1,2}$ is not negligible, 
the pump detuned by $\Omega_2 - \delta$ from the microwave mode $\mwtwo$ 
introduces considerable cross-damping (i.e. resolved sideband cooling) for the fundamental mode.
This cross-damping, measured separately to be 
$\Gammacross \approx 2\pi \product 20$ kHz at the relevant pump powers,
widens the bandwidth of nonreciprocal behaviour by over two orders of magnitude and effectively cools the mechanical oscillator. 
It also acts as loss in the frequency conversion process
and therefore effectively lowers the cooperativities
to $(\mathcal{C}_{11}, \mathcal{C}_{21})\approx (0.78, 0.68)$.
This lowered cooperativity accounts for the overall $\sim$10 dB loss in the forward direction. 
This limitation can be overcome in a future design by increasing 
the sideband resolution with decreased $\kapeye$
or utilising the fundamental modes of two distinct mechanical elements
with similar coupling strengths.
To compare the experiment to theory we use a model 
that takes into account the cross-damping and 
an increased effective mechanical dissipation of the fundamental mode.
The model is compared to the experimental data in
\cref{fig:figure3}E, 
showing good qualitative agreement. 

From a technological standpoint, it does not suffice for an isolator
to have the required transmission properties;
since its purpose is to protect the input from any noise propagating in the backward direction,
the isolator's own noise emission is relevant.
We therefore return to the theoretical description of the ideal symmetric case
and derive the noise properties expected from the device,
in the limit of overcoupled cavities 
($\kappa_{\mathrm{ex},i} \approx \kappa_i$).
In the forward direction and on resonance,
the emitted noise amounts to
$
N_\mathrm{fw}(0)
=
1/2+
(\bar n_{\mathrm{m},1} + \bar n_{\mathrm{m},2}) / (4 \coop)
$,
where $\bar n_{\mathrm{m},j}$
is the thermal occupation of each mechanical mode.
In the limit of low insertion loss and large cooperativity,
the added noise becomes negligible in the forward direction.
More relevant for the purpose of using an isolator
to protect sensitive quantum apparatus is 
the noise emitted in the backward direction,
given by
$
N_\mathrm{bw}(0)
=
1/2 + 
(\bar n_{\mathrm{m},1} + \bar n_{\mathrm{m},2}) / 2 
$.
Here the noise is directly commensurate with the occupation of the mechanics
which can be of hundreds of quanta even at cryogenic millikelvin temperatures, due to
the low mechanical frequencies. 
This is a direct consequence of isolation without reflection, 
since it prevents fluctuations from either cavity to emerge in the backward direction. 
In order to preserve the commutation relations of the bosonic output modes, the fluctuations consequently have to originate from the mechanical modes.
A practical low-noise design
therefore requires a scheme to externally cool the mechanical modes, e.g. via sideband cooling using an additional auxiliary microwave mode.

The origin of this noise asymmetry can be understood as noise interference.
The thermal fluctuations of one mechanical oscillator are
converted to microwave noise in each cavity through two paths,
illustrated in \cref{fig:figure4}A,~D:
a direct (orange) and an indirect (yellow) link.
Each pathway, on its own and with the same coupling strength, 
would result in symmetric noise 
that decreases in magnitude with increasing cooperativity.
When both are present, however, the noise interferes with itself
differently in each direction (cf.\ SI).
In the forward direction, the noise interferes destructively 
(\cref{fig:figure4}B) leading to low added noise,
but in the backward direction a sharp interference peak arises
(\cref{fig:figure4}E) with finite noise in the nonreciprocal bandwidth
even in the high-cooperativity limit.
In an intuitive picture, the circuit acts as a circulator
that routes noise from the output port to the mechanical thermal bath
and in turn the mechanical noise to the input port.
We demonstrate experimentally the noise asymmetry
by detecting the output spectra at each microwave mode 
while the device isolates the mode $\mwone$ from $\mwtwo$
by more than 25 dB
(\cref{fig:figure4}C,~F).
The cooperativities are here set to 
$(\coop_{11}, \coop_{21}, \coop_{12}, \coop_{22})
=
(20.0, 14.2, 106, 89)$
with a cross-damping 
$\Gammacross \approx 2\pi \product 2.6$ kHz,
in order to optimise the circuit for a lower insertion loss
and increase the noise visibility.
As there is additional cooling from the off-resonant pump on mode $\meone$,
we expect noise from $\metwo$ to dominate.
There exists a way to circumvent the mechanical noise entirely:
introducing one extra microwave mode $\mwthree$,
we can realise a circulator, where instead of mechanical fluctuations, the fluctuations from the third microwave mode emerge in the backward direction.
The scheme is illustrated in \cref{fig:figure5}A.
As before, the two mechanical modes are used to create two 
interfering pathways, now between the three microwave cavities.
Since there are now two independent loops, two phases matter;
we choose the phases associated to the couplings $\goo$ and $\gto$
and set them respectively to 
$\phi_1 = 2\pi/3$ and $\phi_2 = -2\pi/3$.
With the mechanical detunings set to 
$\delta_i = 
\frac{\sqrt{3}}{2} (\coop + \frac{1}{3})
\Gammaeye$,
the system then becomes a circulator that routes the input of 
port $\mwone$ to $\mwtwo$, $\mwtwo$ to $\mwthree$ and so on.
Critically and in contrast to above, counter-propagating signals are not dissipated in the mechanical oscillators, 
but directed to the other port, with two advantages.
First, the bandwidth of nonreciprocity is not limited to the 
mechanical dissipation rate, but instead increases with $\coop$
until reaching the ultimate limit given by the cavity linewidth
(see \cref{fig:figure5}B and C).
Second, the mechanical noise emission is symmetrically spread between 
the three modes, and over the wide conversion bandwidth
(see \cref{fig:figure5}D and E).
In the large cooperativity limit, the nonreciprocal process 
becomes quantum limited, irrespective of the temperature
of the mechanical thermal baths.

In conclusion, we described and experimentally demonstrated a new scheme for reconfigurable nonreciprocal transmission in the microwave domain using a superconducting optomechanical circuit.
This scheme is based purely on optomechanical couplings, 
thus
it alleviates the need for coherent microwave cavity-cavity 
(or direct phonon-phonon) interactions, 
and significantly facilitates the experimental realisation, 
in contrast to recently used approaches of optomechanical nonreciprocity 
in the optical domain~\cite{fang_generalized_2017}.
Nonreciprocity arises due to interference in the two mechanical modes,
 which mediate the microwave cavity-cavity coupling. 
This interference also manifests itself in the asymmetric noise output of the circuit.
This scheme can be readily extended to implement quantum-limited phase-preserving and phase-sensitive directional amplifiers~\cite{malz_quantum-limited_2017}.
Moreover, an additional microwave mode enables quantum-limited microwave circulators on-chip with large bandwidth, 
limited only by the energy decay rate of the \emph{microwave} modes.
Finally, the presented scheme can be generalised to an array, 
and thus can form the basis to create topological phases of light and sound~\cite{peano_topological_2015} 
or topologically protected chiral amplifying states~\cite{peano_topological_2016-1} in arrays of electromechanical circuits, 
without requiring cavity-cavity or phonon-phonon mode hopping interactions. 

\begin{acknowledgments}
This work was supported by the SNF, 
the NCCR Quantum Science and Technology (QSIT),  
and the EU Horizon 2020 research
and innovation programme under grant agreement No 732894
(FET Proactive HOT). DM acknowledges support by the UK
Engineering and Physical Sciences Research Council (EPSRC)
under Grant No. EP/M506485/1.
TJK acknowledges financial support from an ERC AdG (QuREM).
AN holds a University Research Fellowship from the Royal Society and acknowledges support from the Winton Programme for the Physics of Sustainability. 
AK holds INSPIRE scholarship from the Department of Science and Technology, India.
All samples were fabricated in the Center of Micro\-Nano\-Technology (CMi) at EPFL.
\end{acknowledgments}

\bibliography{bibliography}

\clearpage

\appendix
\onecolumngrid

\renewcommand\thefigure{\thesection.\arabic{figure}}    
\begin{center}
	\Large \textbf{Supplementary Information}
\end{center}

\section{Theoretical background: Hamiltonian, scattering matrix, nonreciprocity}
\label{SI:sec:theory1}
\setcounter{figure}{0}   

In this section we derive the effective Hamiltonian relevant for our system, calculate the input-output scattering matrix for the electromagnetic modes, and discuss the conditions for obtaining nonreciprocal microwave transmission.

We consider two mechanical degrees of freedom whose positions parametrically modulate the frequencies of two electromagnetic modes via radiation-pressure coupling \cite{cavity_optomechanics_RMP}. The Hamiltonian describing this situation is given by ($\hbar = 1$)
\begin{equation}
\oph
= 
\sum_{i=1}^{2}
\left(
\w_{c,i}\opa_{i}^{\dg}\opa_{i} 
+
\W_{i}\opb_{i}^{\dg}\opb_{i} 
\right)
+
\oph_{\mrm{int}}
+
\oph_{\mrm{drive}},
\label{initH}
\end{equation}
where $\opa_{1}$ and $\opa_{2}$ are the annihilation operators associated with the two electromagnetic modes with frequencies $\w_{c,1}$ and $\w_{c,2}$, and $\opb_{1}$ and $\opb_{2}$ are those for the two mechanical modes with mechanical frequencies $\W_{1}$ and $\Omega_2$, respectively.
Radiation-pressure coupling between the microwave and mechanical modes is described by the interaction Hamiltonian \cite{cavity_optomechanics_RMP} 
\begin{equation}
\oph_{\mathrm{int}}
=
-\sum_{j=1}^{2}
\sum_{k=1}^{2}
g_{0,jk}\;\opa^{\dg}_{j}\opa_{j} 
(\opb_{k} + \opb_{k}^{\dg}),
\label{intH}
\end{equation}
with $g_{0,jk}$ the vacuum optomechanical coupling strength between electromagnetic mode $j$ and mechanical mode $k$
and where we neglect cross coupling terms $\propto \opa^{\dg}_{i}\opa_{j} $, which is a good approximation for spectrally distinct modes $|\w_{c,1} -\w_{c,2} | \gg \Omega_{i}$ \cite{Law1995,Dobrindt2010}.

In the experiment, both cavity modes are driven with two microwave tones each. These four tones are close to the lower mechanical sidebands, but the ones driving the mechanical sidebands at frequency $\Omega_1$ are slightly detuned to the red, whereas the ones driving the sidebands at frequency $\Omega_2$ are slightly detuned to the blue from the lower sideband.
That is, the detuning of the four drives are $\Delta_{jk} = \w_{jk} - \w_{c,j}$ with $\Delta_{11} = \Delta_{21} = - \W_{1} - \delta$ and $\Delta_{12} = \Delta_{22} =  - \W_{2} +\delta$.

We separate mean and fluctuations in the microwave fields and move to a frame rotating at the cavity frequencies
\begin{equation}
\opa_{j} 
= 
e^{-i \w_{c,j} t}
\left(
\left(\delta \opa_{j}\right)
+
\sum_{k=1}^{2}
\alpha_{jk}\;
e^{-i \Delta_{jk} t}
\right)
\label{rot_a}
\end{equation}
where $\alpha_{jk}$ is the coherent state amplitude due to the microwave drive with detuning $\Delta_{jk}$ with $j,k = 1,2$ and $\left(\delta \opa_{j} \right)$ describe the fluctuations of the two microwave modes $j=1,2$.
We then linearize the Hamiltonian by approximating
\begin{equation}
\opa_{j}^{\dg}\opa_{j}
\approx
\left(\delta \opa_{j}^\dagger \right)
\left(
\sum_{k=1}^{2}
\alpha_{jk}\;
e^{-i \Delta_{jk} t}
\right)
+
\mrm{H.c.}
\label{lin_a}
\end{equation}

To obtain a time-independent Hamiltonian we will assume that the system is in the resolved-sideband limit with respect to both mechanical modes, i.e.~$\Omega_1,\Omega_2 \gg \kappa_1,\kappa_2$, and that the two mechanical modes are well separated in frequency, i.e.~$|\Omega_{1}-\Omega_{2}| \gg \Gamma_{m,1}, \Gamma_{m,2}$.
Moving into a rotating frame with respect to the free evolution of the microwave modes, and keeping only non-rotating terms, we obtain the effective Hamiltonian describing our system, which is given as equation (1) in the main manuscript,
\begin{equation}
H = 
-\delta  \, \meone^\dagger \meone
+
\delta  \, \metwo^\dagger \metwo
+
\goo  ( \mwone \meone^\dagger + \mwone^\dagger \meone )
+
\gto  ( \mwtwo \meone^\dagger + \mwtwo^\dagger \meone )
+
\got  ( \mwone \metwo^\dagger + \mwone^\dagger \metwo )
+
\gtt  (  e^{i\phi}\mwtwo  \metwo^\dagger + e^{-i\phi}\mwtwo^\dagger \metwo).
\label{lin_Ham}
\end{equation}
Here, $g_{jk} = g_{0,jk} |\alpha_{jk}|$ are the optomechanical coupling strengths enhanced by the mean intracavity photon numbers $n_{jk} = |\alpha_{jk}|^2$ due to the drive at frequency $\omega_{jk}$ and where we have renamed $\left(\delta \opa_{j} \right) \rightarrow \opa_{j}$ for notational convenience.
Without loss of generality, the phase of all but one coupling constant $g_{jk}$ can be chosen real. Here, we take all of them real and write out the phase $\phi$ explicitly which is varied in our experiment.

From the Hamiltonian (\ref{lin_Ham}) we derive the equations of motion
for our system which can be written in matrix form as \cite{gardiner_input_1985,quantum_noise_RMP}
\begin{equation}
\dot{\mathbf{u}} = \textbf{\texttt{M}} \, \mathbf{u}
+ \textbf{\texttt{L}} \, \mathbf{u}_\textrm{in}
\label{EOM}
\end{equation}
with
$\mathbf{u} = (\hat{a}_1, \hat{a}_2, \hat{b}_1, \hat{b}_2)^T$,
$\mathbf{u}_\textrm{in} = (\hat{a}_\textrm{1,in}, \hat{a}_\textrm{2,in}, \hat{a}_\textrm{1,in}^{(0)}, \hat{a}_\textrm{2,in}^{(0)},\hat{b}_\textrm{1,in}, \hat{b}_\textrm{2,in})^T$
and
$\mathbf{u}_\textrm{out} = (\hat{a}_\textrm{1,out}, \hat{a}_\textrm{2,out}, \hat{a}_\textrm{1,out}^{(0)}, \hat{a}_\textrm{2,out}^{(0)},\hat{b}_\textrm{1,out}, \hat{b}_\textrm{2,out})^T$,
where $\hat{a}_{i,\textrm{in/out}}$ are the input-output modes of the external microwave feedline and $\hat{a}_{i,\textrm{in/out}}^{(0)}$ are those corresponding to internal dissipation.

The matrix $\textbf{\texttt{M}}$ reads
\begin{equation}
\textbf{\texttt{M}} =
\left(
\begin{matrix}
-\frac{\kappa_1}{2} & 0 & -ig_{11} & -ig_{12} \\
0 & -\frac{\kappa_2}{2} & -ig_{21} & -ig_{22}e^{-i\phi} \\
-ig_{11} & -ig_{21} & +i\delta-\frac{\Gamma_\text{m,1}}{2} & 0\\
-ig_{12} & -ig_{22}e^{+i\phi} & 0 & -i\delta-\frac{\Gamma_\text{m,2}}{2}
\end{matrix}
\right)
\end{equation}
where the cavity dissipation rates are the sum of external and internal dissipation rates, i.e.~$\kappa_1 = \kappa_\text{ex,1}+\kappa_\text{0,1}$
and $\kappa_2 = \kappa_\text{ex,2}+\kappa_\text{0,2}$,
and the matrix $\textbf{\texttt{L}}$ reads
\begin{equation}
\textbf{\texttt{L}} =
\left(
\begin{matrix}
\sqrt{\kappa_\text{ex,1}} & 0 & \sqrt{\kappa_\text{0,1}} & 0 & 0 & 0 \\
0 & \sqrt{\kappa_\text{ex,2}} & 0 & \sqrt{\kappa_\text{0,2}} & 0 & 0 \\
0 & 0 & 0 & 0 & \sqrt{\Gamma_\text{m,1}} & 0\\
0 & 0 & 0 & 0 & 0 & \sqrt{\Gamma_\text{m,2}}
\end{matrix}
\right).
\end{equation}

Using the input-output relations for a one-sided cavity \cite{gardiner_input_1985,quantum_noise_RMP}
\begin{equation}
\mathbf{u}_{\textrm{out}} = \mathbf{u}_{\textrm{in}}-\textbf{\texttt{L}}^T \, \mathbf{u}
\label{inputoutput}
\end{equation}
we can solve the input-output problem in the Fourier domain
\begin{equation}
\mathbf{u}_{\textrm{out}}(\omega)=S(\omega)\, \mathbf{u}_{\textrm{in}}(\omega)
\end{equation}
with the scattering matrix
\begin{equation}
S(\omega) = \mathbb{1}_{6\times6} + \textbf{\texttt{L}}^T \, [+i \omega \mathbb{1}_{4\times4} + \textbf{\texttt{M}}]^{-1} \, \textbf{\texttt{L}}.
\label{sca_mat}
\end{equation}

Eliminating the mechanical degrees of freedom from the equations of motion (\ref{EOM}) we obtain
\begin{align}
&\left(
\begin{matrix}
\frac{\kappa_1}{2}-i\omega + g_{11}^2 \chi_1(\omega) + g_{12}^2 \chi_2(\omega)
& g_{11} \chi_1(\omega) g_{21} + g_{12} \chi_2(\omega) g_{22} e^{+i\phi}\\
g_{11} \chi_1(\omega) g_{21} + g_{12} \chi_2(\omega) g_{22} e^{-i\phi}
& \frac{\kappa_2}{2}-i\omega + g_{21}^2 \chi_1(\omega) + g_{22}^2 \chi_2(\omega)
\end{matrix}
\right)
\left(
\begin{matrix}
\hat{a}_1 \\
\hat{a}_2
\end{matrix}
\right) \nonumber \\
&=
\left(
\begin{matrix}
\sqrt{\kappa_{\text{ex},1}}\hat{a}_\text{in,1}
+ \sqrt{\kappa_{0,1}}\hat{a}_\text{in,1}^{(0)}
- i g_{11} \chi_1(\omega) \sqrt{\Gamma_{\text{m},1}} \hat{b}_{1,\text{in}}
- i g_{12} \chi_2(\omega) \sqrt{\Gamma_{\text{m},2}} \hat{b}_{2,\text{in}}\\
\sqrt{\kappa_{\text{ex},2}}\hat{a}_\text{in,2}
+ \sqrt{\kappa_{0,2}}\hat{a}_\text{in,2}^{(0)}
- i g_{21} \chi_1(\omega) \sqrt{\Gamma_{\text{m},1}} \hat{b}_{1,\text{in}}
- i g_{22} \chi_2(\omega) e^{-i\phi} \sqrt{\Gamma_{\text{m},2}} \hat{b}_{2,\text{in}}
\end{matrix}
\right)
\label{intermed_step}
\end{align}
where we introduced the mechanical susceptibilities
$\chi_1^{-1} (\omega) = \Gammaone/2 - i \left( \delta + \omega \right)$
and
$\chi_2^{-1} (\omega) = \Gammatwo/2 + i \left( \delta - \omega \right)$.
Inverting the matrix in equation (\ref{intermed_step}) and exploiting the input-output relation (\ref{inputoutput}), we obtain equation (3) of the main manuscript
\begin{equation}
\frac{S_{12}(\omega)}{S_{21}(\omega)}
= 
\frac{
	g_{11}
	\chi_{1}
	(
	\w
	)
	g_{21}
	+
	g_{12}
	\chi_{2}
	(
	\w
	)
	g_{22}
	e^{+i\phi}
}{
	g_{11}
	\chi_{1}
	(
	\w
	)
	g_{21}
	+
	g_{12}
	\chi_{2}
	(
	\w
	)
	g_{22}
	e^{-i\phi}
}.
\label{Sratio}
\end{equation}
Note that the expressions in the nominator and denominator in (\ref{Sratio}) are equal to the matrix elements coupling the two electromagnetic modes in (\ref{intermed_step}) which are the sum of the two (complex) amplitudes for the two dissipative optomechanical pathways.
Equation (\ref{Sratio}) 
 is used to generate Fig.\ 3 E of the main text,
with all the parameters ($\Gammajay, g_{ij}$) independently measured.

For identical mechanical decay rates 
$\Gamma_{\mathrm{m},1} = \Gamma_{\mathrm{m},2} = \Gammasym$ 
and identical cooperativities 
$\coop = \coop_{ij} = \frac{4g_{ij}^2}{\kappa_i\Gamma_{m,j}}$ 
we find that the transmission $2\rightarrow 1$ vanishes on resonance $\omega=0$, i.e.~$S_{12} = 0$, if
\begin{equation}
\frac{\Gammasym}{2\delta} = \tan \frac{\phi}{2}.
\label{ph_cond}
\end{equation}
For a given $\delta$,
maximal transmission in the opposite direction $1\rightarrow 2$ is then obtained for 
$\coop = \frac{1}{2} + \frac{2 \delta^2}{\Gammasym^2}$ 
and given by
\begin{equation}
  |S_{21}|^{2}_\text{max}
  = \frac{\kappa_{\text{ex},1}\kappa_{\text{ex},2}}{\kappa_1\kappa_2}\frac{4 \delta^2}{\Gammasym^2 + 4 \delta^2}
  = \frac{\kappa_{\text{ex},1}\kappa_{\text{ex},2}}{\kappa_1\kappa_2}\left(1 - \frac{1}{2 \coop}\right). 
  \label{maxtransm}
\end{equation}
We see that for $\delta \gg \Gammasym$ the optimal cooperativity $\coop \rightarrow \infty$ and $|S_{21}(0)|^{2} \rightarrow 1$. 
Thus, we see that in this limit the electromagnetic scattering matrix of our system becomes that of an ideal isolator, i.e.~$S_{11}=S_{12}=S_{22}=0$ and $|S_{21}| = 1$.

\begin{figure}
\centering
\includegraphics
{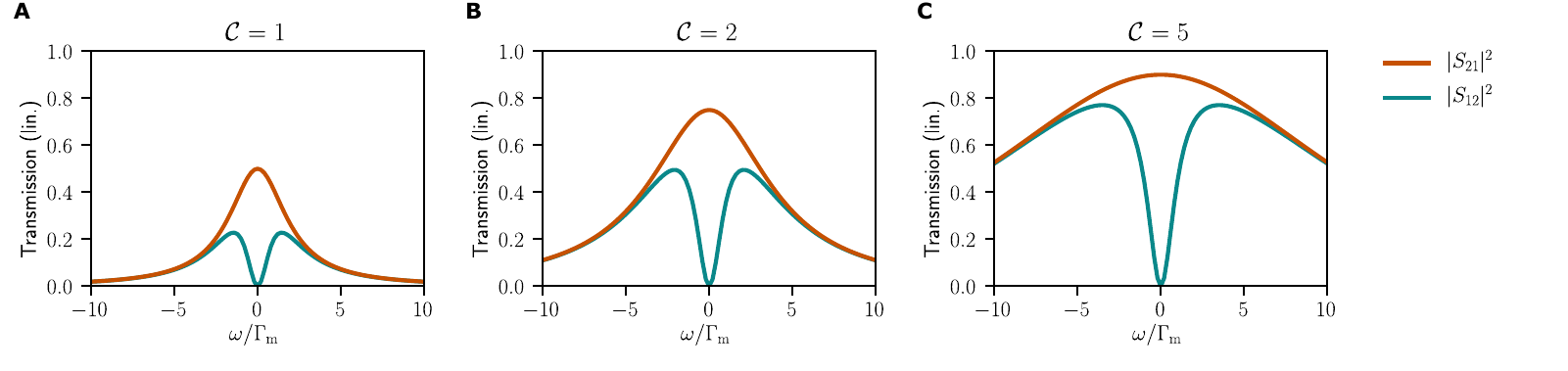}
\caption{
  Microwave transmission of the nonreciprocal electromechanical device in each direction for different values of 
  the cooperativity $\coop$,
  derived from \cref{sca_mat}
  for the case of symmetric mechanical modes 
  ($\Gammaone = \Gammatwo = \Gammasym$).
  The detuning $\delta$ and the phase $\phi$ are set for 
  maximal transmission according to \cref{maxtransm}.
  As the cooperativity is increased, the overall bandwidth
  of the frequency conversion increases to $\Gamma_\mathrm{eff}$, however
  the bandwidth of nonreciprocal transmission stays
  constant and is on the order of 
  the intrinsic mechanical damping rate $\Gammasym$.
  This illustrates the fact that the intrinsic dissipation 
  of the mechanical oscillator is the underlying resource for 
  the nonreciprocity.
}
\label{fig:NRbandwidth}
\end{figure}

The full scattering matrix $S_{ij}$ of (\ref{sca_mat})
is used 
in \cref{fig:NRbandwidth}
to show optimal transmission in each direction for the symmetric case,
with different values of the cooperativity.
As the cooperativity increases, the overall bandwidth of conversion
increases to $\Gamma_\mathrm{eff}$,
but the nonreciprocal bandwidth stays constant.
This can be seen in the ratio 
$S_{12}(\omega)/S_{21}(\omega)$
in (\ref{Sratio})
that depends only on the bare mechanical susceptibilities
$\chi_1(\omega)$ and $\chi_2(\omega)$.

For unequal decay rates $\Gamma_{\mathrm{m},1} \not = \Gamma_{\mathrm{m},2}$, but equal effective decay rates of the mechanical modes $\Gamma_{\eff,j} = \Gamma_{\text{m},j}(1 + \coop_{1j} + \coop_{2j}$), nonreciprocity is obtained for $\frac{\Gamma_{+}}{2\delta} = \tan \frac{\phi}{2}$ off-resonance at a frequency $\omega = \frac{\Gamma_{+}\Gamma_{-}}{4\delta}$ where $\Gamma_{\pm} = \frac{1}{2}\left(\Gamma_{\mathrm{m},1} \pm \Gamma_{\mathrm{m},2}\right)$.
For unequal decay rates $\Gamma_{\mathrm{m},1} \not = \Gamma_{\mathrm{m},2}$, but matched cooperativities $\coop_{jk} = \coop$, we find nonreciprocity for $\frac{\Gamma_{+}}{2\delta} = \tan \frac{\phi}{2}$, but at $\omega = - \frac{\Gamma_{-}\delta}{\Gamma_{+}}$.

\section{Theoretical background: Noise analysis of the device}
\label{SI:sec:theory2}

In this section we analyse the noise properties of the nonreciprocal electromechanical device. We assume the bosonic input noise operators obey
\begin{align}
\label{noise_first}
\langle \hat{a}_\textrm{1,in}(t) \hat{a}_\textrm{1,in}^\dagger(t') \rangle
&= \delta(t-t')\\
\langle \hat{a}_\textrm{2,in}(t) \hat{a}_\textrm{2,in}^\dagger(t') \rangle
&= \delta(t-t')\\
\langle \hat{a}_\textrm{1,in}^{(0)}(t) \hat{a}_\textrm{1,in}^{(0)\dagger}(t') \rangle
&= \delta(t-t')\\
\langle \hat{a}_\textrm{2,in}^{(0)}(t) \hat{a}_\textrm{2,in}^{(0)\dagger}(t') \rangle
&= \delta(t-t')\\
\langle \hat{b}_\textrm{1,in}(t) \hat{b}_\textrm{1,in}^\dagger(t') \rangle
&= (\bar{n}_\text{m,1}+1)\delta(t-t')\\
\langle \hat{b}_\textrm{2,in}(t) \hat{b}_\textrm{2,in}^\dagger(t') \rangle
&= (\bar{n}_\text{m,2}+1)\delta(t-t'),
\label{noise_last}
\end{align}
i.e.~the baths of the microwave modes are assumed to be at zero temperature whereas the mechanical modes have a finite thermal occupation $\bar{n}_{\text{m},1}$ and $\bar{n}_{\text{m},2}$, respectively.

The symmetrised output noise spectra \cite{quantum_noise_RMP} are determined by the scattering matrix of the device (\ref{sca_mat}) as well as the noise properties of the microwave and mechanical baths (\ref{noise_first})-(\ref{noise_last}). Explicitly, we find that the cavity output spectra are given by
\begin{equation}
	\begin{aligned}
		\bar{S}_\text{1,out}(\omega)
		&= \frac{1}{2} \int_{-\infty}^{\infty} \!\!\! dt \, e^{i\omega t}
		\langle \hat{a}_\text{1,out}^\dagger(t) \hat{a}_\text{1,out}(0)
		+ \hat{a}_\text{1,out}(0) \hat{a}_\text{1,out}^\dagger(t)\rangle\\
		&=\tfrac{1}{2} \left[ |S_{11}(-\omega)|^2 +|S_{12}(-\omega)|^2 +|S_{13}(-\omega)|^2  +|S_{14}(-\omega)|^2 \right]
		+|S_{15}(-\omega)|^2 (\bar{n}_\text{m,1}+\tfrac{1}{2})+|S_{16}(-\omega)|^2 (\bar{n}_\text{m,2}+\tfrac{1}{2})
	\end{aligned}
\end{equation}
and
\begin{equation}
	\begin{aligned}
		\bar{S}_\text{2,out}(\omega)
		&= \frac{1}{2} \int_{-\infty}^{\infty} \!\!\! dt \, e^{i\omega t}
		\langle \hat{a}_\text{2,out}^\dagger(t) \hat{a}_\text{2,out}(0)
		+ \hat{a}_\text{2,out}(0) \hat{a}_\text{2,out}^\dagger(t)\rangle\\
		&=\tfrac{1}{2} \left[ |S_{21}(-\omega)|^2 +|S_{22}(-\omega)|^2 +|S_{23}(-\omega)|^2  +|S_{24}(-\omega)|^2 \right]
		+|S_{25}(-\omega)|^2 (\bar{n}_\text{m,1}+\tfrac{1}{2})+|S_{26}(-\omega)|^2 (\bar{n}_\text{m,2}+\tfrac{1}{2}).
	\end{aligned}
\end{equation}

In the limit of overcoupled cavities $\kappa_{\text{ex},i} \approx \kappa_{i}$ and for the optimal phase $\phi$ and detuning $\delta$, the noise emitted in the backward direction $2\rightarrow 1$ on resonance $\omega=0$ is
\begin{equation}
	\begin{aligned}
		N_\text{bw} = \bar{S}_\text{1,out}(0)
		&= |S_{11}|^2 \cdot \frac{1}{2}
		+ |S_{12}|^2 \cdot \frac{1}{2}
		+ |S_{15}|^2 \cdot \left(\bar{n}_\text{m,1} + \frac{1}{2}\right)
		+ |S_{16}|^2 \cdot \left(\bar{n}_\text{m,2} + \frac{1}{2}\right)\\
		&= 
		0 \cdot \frac{1}{2}
		+ 0 \cdot \frac{1}{2}
		+ \frac{1}{2} \cdot \left(\bar{n}_\text{m,1} + \frac{1}{2}\right)
		+ \frac{1}{2} \cdot \left(\bar{n}_\text{m,2} + \frac{1}{2}\right)\\
		&= \frac{1}{2} + \frac{\bar{n}_\text{m,1} + \bar{n}_\text{m,2}}{2},
	\end{aligned}
\end{equation}
i.e.~in the backward direction the noise of the device is dominated by the noise emitted by the mechanical oscillators.

The noise emitted in the forward direction $1\rightarrow 2$ on resonance $\omega=0$ is
\begin{equation}
	\begin{aligned}
		N_\text{fw} = \bar{S}_\text{2,out}(0)
		&= |S_{21}|^2 \cdot \frac{1}{2} + |S_{22}|^2 \cdot \frac{1}{2}
		+ |S_{25}|^2 \cdot \left(\bar{n}_\text{m,1} + \frac{1}{2}\right)
		+ |S_{26}|^2 \cdot \left(\bar{n}_\text{m,2} + \frac{1}{2}\right)\\
		&= \left(1-\frac{1}{2\mathcal{C}}\right) \cdot \frac{1}{2}
		+ 0 \cdot \frac{1}{2}
		+ \frac{1}{4\mathcal{C}} \cdot \left(\bar{n}_\text{m,1} + \frac{1}{2}\right)
		+ \frac{1}{4\mathcal{C}} \cdot \left(\bar{n}_\text{m,2} + \frac{1}{2}\right)\\
		&= \frac{1}{2} + \frac{\bar{n}_\text{m,1} + \bar{n}_\text{m,2}}{4 \mathcal{C}},
	\end{aligned}
\end{equation}
i.e.~in the forward direction the noise contribution from the mechanical oscillators vanishes at large cooperativity $\mathcal{C}\gg 1$. 
Therefore, intriguingly, the noise emitted on resonance by the nonreciprocal device is not symmetric in the forward and backward directions.

\section{Theoretical background: Noise interference as origin of asymmetric noise emission}
In the previous section we concluded that the circuit emits more noise in the backward direction as compared to the forward direction.
This is also corroborated by the experimental data, shown in Fig.~4 in the main text. 
In the following, in order to understand the different noise performance in the forward and backward direction, we consider two different points of view.
First, we derive the scattering amplitude from one mechanical resonator to one cavity, eliminating the other two modes.
In this picture, the imbalance can be understood as an interference of the two paths the noise can take in the circuit, analogously to the interference in the microwave transmission.
Second, we eliminate the mechanical resonators, but taking their input noise into account. 
This leads to the same scattering matrix for the microwaves as discussed in the main text,
but the mechanical noise appears as additional, effective noise input operators for the cavities.
In the second formulation we can therefore use our knowledge of the microwave scattering matrix to deduce properties of the noise scattering.

Let us first consider the scattering from a mechanical resonator to cavities 1 and 2. 
Since in the experiment mechanical resonator 1 is strongly cross-damped due to off-resonant couplings, the noise emitted stems almost exclusively from resonator 2.
If we are interested in the noise scattering from mechanical resonator 2 to cavity 2, we can eliminate the other two modes and drop their input noise.
In frequency space, their equations of motion are
\begin{equation}
	\mat{\chi_{c,1}^{-1}(\omega)&-ig_{11}\\-ig_{11}^*&\chi_{1}^{-1}(\omega)}\mat{\hat a_1\\\hat b_1}=\mat{ig_{12}\hat b_2\\ig_{21}^*\hat a_2}+\text{noises}.
\end{equation}
We drop the noise terms and solve for $\hat a_1,\hat b_1$ 
\begin{equation}
  \begin{aligned}
	\mat{\hat a_1(\omega)\\\hat b_1(\omega)}&=\frac{1}{\chi_{1}^{-1}(\omega)\chi_{c,1}^{-1}(\omega)+|g_{11}|^2}
	\mat{\chi_{1}^{-1}(\omega)&ig_{11}\\ ig_{11}^*&\chi_{c,1}^{-1}(\omega)}
	\mat{ig_{12}\hat b_2(\omega)\\ig_{21}^*\hat a_2(\omega)}\\
	&\equiv\chi_{\hat a_1\hat b_1}(\omega)\mat{ig_{12}&\\&ig_{21}^*}\mat{\hat b_2(\omega)\\\hat a_2(\omega)},
  \end{aligned}
  \label{eq:a1b1}
\end{equation}
where we defined the cavity susceptibility $\chi_{c,i}^{-1}(\omega)=\kappa_i/2-i\omega$
and the susceptibility of the coupled system of modes $\hat a_1,\hat b_1$, $\chi_{\hat a_1\hat b_1}(\omega)$.
We turn to the other two modes, the ones that we are actually interested in.
For those, we have a similar equation, which can be obtained from interchanging $1\leftrightarrow2$
\begin{equation}
	\mat{\chi_{2}^{-1}(\omega)&-ig_{22}^*\\-ig_{22}&\chi_{c,2}^{-1}(\omega)}\mat{\hat b_2(\omega)\\\hat a_2(\omega)}
	=\mat{ig_{12}^*&\\&ig_{21}}\mat{\hat a_1(\omega)\\\hat b_1(\omega)}+\mat{\sqrt{\Gamma_{\mathrm m,2}}\hat b_{2,\text{in}}(\omega)\\ \sqrt{\kappa_2}\hat a_{2,\text{in}}(\omega)}.
\end{equation}
Eliminating the modes $\hat a_1,\hat b_1$ with \cref{eq:a1b1}, we arrive at
\begin{equation}
  \begin{aligned}
	  \mat{\sqrt{\Gamma_{\mathrm m,2}}\hat b_{2,\text{in}}(\omega)\\ \sqrt{\kappa_2}\hat a_{2,\text{in}}(\omega)}
	&=
	\left[ \mat{\chi_{2}^{-1}(\omega)&-ig_{22}^*\\-ig_{22}&\chi_{c,2}^{-1}(\omega)}
	-\frac{\mat{ig_{12}^*&\\&ig_{21}}
	\mat{\chi_{1}^{-1}(\omega)&ig_{11}\\ig_{11}^*&\chi_{c,1}^{-1}(\omega)}
	\mat{ig_{12}&\\&ig_{21}^*}
	}{\chi_{1}^{-1}(\omega)\chi_{c,1}^{-1}(\omega)+|g_{11}|^2}
	\right]\mat{\hat b_2(\omega)\\\hat a_2(\omega)}\\
	&\equiv \left[ \chi_{\hat b_2\hat a_2}^{-1}(\omega)-\mat{ig_{12}^*&\\&ig_{21}}\chi_{\hat a_1\hat b_1}(\omega)\mat{ig_{12}&\\&ig_{21}^*} \right]
	\mat{\hat b_2(\omega)\\\hat a_2(\omega)}.
  \end{aligned}
\end{equation}
In the second line, we have formulated the equation in terms of the susceptibilities of the two subsystems $(\hat a_1,\hat b_1)$ and $(\hat a_2,\hat b_2)$.
This equation is a bit complicated, but we note that the coupling between $\hat a_2$ and $\hat b_2$ is
\begin{equation}
  	iT_{\hat a_2\hat b_2}(\omega)=-ig_{22}
  	\left[ 1-e^{-i\phi_p}\frac{\cc_{12}\cc_{21}/(\cc_{22}\cc_{11})}{1+\left( \chi_{c,1}(\omega)\chi_{1}(\omega)|g_{11}|^2 \right)^{-1}} \right].
  	\label{eq:T22}
\end{equation}
Analogously, changing the indices referring to the cavity, we obtain the coupling between $\hat a_1$ and $\hat b_2$
\begin{equation}
  	iT_{\hat a_1\hat b_2}(\omega)=-ig_{12}
  	\left[ 1-e^{+i\phi_p}\frac{\cc_{11}\cc_{22}/(\cc_{12}\cc_{21})}{1+\left( \chi_{c,2}(\omega)\chi_{1}(\omega)|g_{21}|^2 \right)^{-1}} \right],
	\label{eq:T12}
\end{equation}
The coupling phase $\phi_p$ appears as the relative phase between indirect and direct coupling path, as for the microwave signal transmission.
\Cref{eq:T22,eq:T12} demonstrate that the transmission of noise from the mechanical resonators to the microwave cavities is subject to interference, which ultimately leads to the difference in noise emitted in the forward versus the backward direction. 

In a second picture, we can also understand the mechanical noise interference in terms of the nonreciprocity in the scattering matrix for the microwave modes.
In order to do so, we solve the equations of motion for the mechanical resonators (given in \cref{EOM}), which leads to
\begin{equation}
  	\hat b_j(\omega)=\chi_{j}(\omega)\left[ i\sum_ig_{ij}^*\hat a_i(\omega)+\sqrt{\Gamma_{\mathrm m,j}}\hat b_{j,\text{in}}(\omega) \right].
  \label{eq:b elimination}
\end{equation}
We obtain equations that only relate the cavities
\begin{equation}
	\mat{\chi_{c,1}^{-1}(\omega)+iT_{11}(\omega)&iT_{12}(\omega)\\iT_{21}(\omega)&\chi_{c,2}^{-1}(\omega)+iT_{22}(\omega)}\mat{\hat a_1(\omega)\\\hat a_2(\omega)}
	=i\mat{g_{11}&g_{12}\\g_{21}&g_{22}}\mat{\sqrt{\Gamma_{\mathrm m,1}}\chi_{1}(\omega)\hat b_{1,\text{in}}(\omega)\\\sqrt{\Gamma_{\mathrm m,2}}\chi_{2}(\omega)\hat b_{2,\text{in}}(\omega)}
	+\mat{\sqrt{\kappa_1}\hat a_{1,\text{in}}(\omega)\\ \sqrt{\kappa_2}\hat a_{2,\text{in}}(\omega)},
\end{equation}
where
\begin{equation}
  	iT_{ij}(\omega)\equiv -i\sum_k\chi_{k}(\omega)g_{ik}g_{jk}^*.
  \label{eq:Tij}
\end{equation}

We can think of mechanical noise as coloured and correlated noise in the optical inputs.
That is, consider the replacement
\begin{equation}
	\mat{\sqrt{\kappa_1}\hat c_{1,\text{in}}(\omega)\\ \sqrt{\kappa_2}\hat c_{2,\text{in}}(\omega)}\equiv
	i\mat{g_{11}&g_{12}\\g_{21}&g_{22}}\mat{\sqrt{\Gamma_{\mathrm m,1}}\chi_{1}(\omega)\hat b_{1,\text{in}}(\omega)\\
	\sqrt{\Gamma_{\mathrm m,2}}\chi_{2}(\omega)\hat b_{2,\text{in}}(\omega)}.
\end{equation}
The effective noise $\hat c_{i,\text{in}}$ is both coloured
$\ev{\hat c_{1,\text{in}}\dagg(\omega)\hat c_{1,\text{in}}(\omega')}\neq \delta(\omega+\omega')\bar n_{\text{1,eff}}$
and correlated $\ev{\hat c_{1,\text{in}}\dagg(\omega) \hat c_{2,\text{in}}(\omega')}\neq0$.

Using the input-output relation $\hat a_{\text{out}}=\hat a_{\text{in}}-\sqrt{\kappa}\hat a$,
the cavity output is given by
\begin{equation}
	\mat{\hat a_{1,\text{out}}(\omega)\\\hat a_{2,\text{out}}(\omega)}
	=S(\omega)\mat{\hat a_{1,\text{in}}(\omega)\\\hat a_{2,\text{in}}(\omega)}
	+[S(\omega)-\id_2]\mat{\hat c_{1,\text{in}}(\omega)\\\hat c_{2,\text{in}}(\omega)},
	\label{eq:cavity output}
\end{equation}
where in the last step we have identified the 2-by-2 optical scattering matrix $S(\omega)$ that relates the cavity inputs to the outputs
$\hat a_{i,\text{out}}(\omega)=\sum_jS_{ij}(\omega)\hat a_{j,\text{in}}(\omega)$.
The fact that \cref{eq:cavity output} contains mechanical noise as well, but can be written entirely in terms of the optical scattering matrix constitutes the central result here.
Since the two effective input noises $\hat c_{i,\text{in}}$ are coloured and correlated, they can interfere.

Most importantly, we can consider what happens when the circuit is impedance matched to the signal and perfectly isolating.
We choose the detunings $\delta_1=\Gamma_{\mathrm m,1}\delta/2,\delta_2=-\Gamma_{\mathrm m,2}\delta/2$,
for some dimensionless parameter $\delta$.
For simplicity, let us choose all cooperativities to be equal $\cc=\cc_{ij}$.
For $\delta^2=2\cc-1$ (impedance matching), the optical scattering matrix of the isolator is (up to some irrelevant phase)
\begin{equation}
	S(0)=\mat{0&0\\ \sqrt{1-1/(2\cc)}&0}\equiv T\mat{0&0\\1&0}.
	\label{eq:ideal Sopt}
\end{equation}
The cavity output on resonance is
\begin{equation}
	\mat{\hat a_{1,\text{out}}\\\hat a_{2,\text{out}}}
	=T\mat{0\\\hat a_{1,\text{in}}}
	-\frac{i}{\sqrt{2}}\cc\mat{e^{i\phi_p}&1\\1-Te^{i\phi_p}&1-T}
	\mat{\hat b_{1,\text{in}}(0)\\\hat b_{2,\text{in}}(0)}.
\end{equation}
As $\cc\to\infty$, $T\to1$ and $\phi_p=\arccos(1-1/\cc)\to0$, such that the second cavity does not receive any noise,
which is due to an interference of $\hat c_{1,\text{in}}$ with $\hat c_{2,\text{in}}$.
In the backward direction, no interference can take place, since cavity 2 is isolated from cavity 1.
As a consequence, the number of noise quanta emerging from cavity 1 on resonance is $N_{\text{bw}}=(\bar n_{\text{m},1}+\bar n_{\mathrm m,2}+1)/2$.

\section{Theoretical background: optomechanical circulator}
We consider three microwave modes (described by their annihilation operators $\hat{a}_1$, $\hat{a}_2$, $\hat{a}_3$) with  resonance frequencies $\omega_{c,1}$, $\omega_{c,2}$, $\omega_{c,3}$ and dissipation rates $\kappa_1$, $\kappa_2$, $\kappa_3$. These three microwave modes are coupled to two mechanical modes (described by the annihilation operators $\hat{b}_1$, $\hat{b}_2$) with resonance frequencies $\Omega_1$, $\Omega_2$ and dissipation rates $\Gamma_{\mathrm{m},1}$ and $\Gamma_{\mathrm{m},2}$. The optomechanical coupling strengths $g_{ij}$ are taken to be real and we write out three phases ($\phi_1$, $\phi_2$, $\phi_3$). Each loop has a relevant phase which is a linear combination of those previous phases. The three cavities are driven with two microwave tones each. These six tones are close to the lower motional sidebands, with detunings of $\Delta_{11}=\Delta_{21}=\Delta_{31}=-\Omega+\delta_1$ and $\Delta_{12}=\Delta_{22}=\Delta_{32}=-\Omega_2+\delta_2$; $\delta_1$ and $\delta_2$ are to be determined. The cooperativities are matched $\mathcal{C}=\mathcal{C}_{ij}=\frac{4g_{ij}}{\kappa_i\Gamma_j}$. \\
The linearised Hamiltonian, describing the system, in a frame rotating with the cavity frequencies and keeping non-rotating terms is given by
\begin{equation}
\begin{aligned}
\hat{H}=&\delta_1\hat{b}_1^\dagger\hat{b}_1+\delta_2\hat{b}_2^\dagger\hat{b}_2\\
&+g_{11}(\hat{a}_1^\dagger\hat{b}_1e^{i \phi_1}+\hat{a}_1\hat{b}_1^\dagger e^{-i \phi_1})+g_{12}(\hat{a}_1^\dagger\hat{b}_2+\hat{a}_1\hat{b}_2^\dagger)\\
&+g_{21}(\hat{a}_2^\dagger\hat{b}_1e^{i \phi_2}+\hat{a}_2\hat{b}_1^\dagger e^{-i \phi_2})+g_{22}(\hat{a}_2^\dagger\hat{b}_2+\hat{a}_2\hat{b}_2^\dagger)\\
&+g_{31}(\hat{a}_3^\dagger\hat{b}_1e^{i \phi_3}+\hat{a}_3\hat{b}_1^\dagger e^{-i \phi_3})+g_{32}(\hat{a}_3^\dagger\hat{b}_2+\hat{a}_3\hat{b}_2^\dagger ).
\end{aligned}
\label{HamCirculator5modes}
\end{equation}
From the Hamiltonian (\ref{HamCirculator5modes}), we derive the equations of motion for our system in their matrix form (\ref{EOM}) with \sloppy $\mathbf{u}=(\hat{a}_1, \hat{a}_2,\hat{a}_3,\hat{b}_1,\hat{b}_2)^T$, $\mathbf{u}_\mathrm{in} =(\hat{a}_\mathrm{1,in}, \hat{a}_\mathrm{2,in}, \hat{a}_\mathrm{3,in}, \hat{a}_\mathrm{1,in}^{(0)}, \hat{a}_\mathrm{2,in}^{(0)}, \hat{a}_\mathrm{3,in}^{(0)},\hat{b}_\mathrm{1,in}, \hat{b}_\mathrm{2,in})^T $ and  $\mathbf{u}_\mathrm{out} = (\hat{a}_\mathrm{1,out}, \hat{a}_\mathrm{2,out}, \hat{a}_\mathrm{3,out},\hat{a}_\mathrm{1,out}^{(0)}, \hat{a}_\mathrm{2,out}^{(0)},\hat{a}_\mathrm{3,out}^{(0)},\hat{b}_\mathrm{1,out}, \hat{b}_\mathrm{2,out})^T$. The $\textbf{\texttt{M}}$ matrix reads 
\begin{equation}
\textbf{\texttt{M}}=
\left(
\begin{array}{ccccc}
 -\frac{\kappa_1}{2} & 0 & 0 & -i g_{11} e^{i \phi_1} & -i g_{12} \\
 0 & -\frac{\kappa_2}{2} & 0 & -i g_{21} e^{i \phi_2} & -i g_{22} \\
 0 & 0 & -\frac{\kappa_3}{2} & -i g_{31} e^{i \phi_3} & -i g_{32} \\
 -i g_{11} e^{-i \phi_1} & -i g_{21} e^{-i \phi_2} & -i g_{31} e^{-i \phi_3} & -\frac{\Gamma_\text{m,1}}{2}+i
   \delta_1 & 0 \\
 -i g_{12} & -i g_{22} & -i g_{32} & 0 & -\frac{\Gamma_\text{m,2}}{2}+i \delta_2 \\
\end{array}
\right),
\end{equation}
where the cavity dissipation rates are the sum of the external and internal dissipation rates, i.e. $\kappa_i=\kappa_{\mathrm{ex},i}+\kappa_{0,i}$, and the $\textbf{\texttt{L}}$ matrix reads
\begin{equation}
\textbf{\texttt{L}}=
\left(
\begin{array}{cccccccc}
 \sqrt{\kappa_\text{ex,1}} & 0 & 0 & \sqrt{\kappa_{0,1}} & 0 & 0 & 0 & 0 \\
 0 & \sqrt{\kappa_\text{ex,2}} & 0 & 0 & \sqrt{\kappa_{0,2}} & 0 & 0 & 0 \\
 0 & 0 & \sqrt{\kappa_\text{ex,3}} & 0 & 0 & \sqrt{\kappa_{0,3}} & 0 & 0 \\
 0 & 0 & 0 & 0 & 0 & 0 & \sqrt{\Gamma_\text{m,1}} & 0 \\
 0 & 0 & 0 & 0 & 0 & 0 & 0 & \sqrt{\Gamma_\text{m,2}} \\
\end{array}
\right).
\end{equation}
Using the input-output relation (\ref{inputoutput}) and the matrix form of the equations of motion (\ref{EOM}), we can compute the scattering matrix $S(\omega)$ similarly to (\ref{sca_mat}).\\
We arbitrarily choose to suppress the propagation in the clockwise direction, i.e. $S_{12}(0)=0$, $S_{23}(0)=0$, and $S_{31}(0)=0$. To enforce this suppression to occur on resonance ($\omega=0$) ,  $\delta_1$ scales with $\Gamma_\text{m,1}$ and $\delta_2$ with $\Gamma_\text{m,2}$, i.e. $\delta_1=\alpha\Gamma_\text{m,1}$ and $\delta_2=\beta\Gamma_\text{m,2}$. \\
The set of equations corresponding to $S_{12}(0)=S_{23}(0)=S_{31}(0)=0$ is given by
\begin{equation}
\begin{aligned}
\begin{cases}
-2i\alpha-2i\beta e^{i(\phi_1-\phi_2)}-\mathcal{C}(1-e^{i(\phi_1-\phi_3)}-e^{i(\phi_3-\phi_2)}+e^{i(\phi_1-\phi_2)})-(1+e^{i(\phi_1-\phi_2)})&=0\\
-2i\alpha-2i\beta e^{i(\phi_2-\phi_3)}-\mathcal{C}(1-e^{i(\phi_2-\phi_1)}-e^{i(\phi_1-\phi_3)}+e^{i(\phi_2-\phi_3)})-(1+e^{i(\phi_2-\phi_3)})&=0\\
-2i\alpha-2i\beta e^{i(\phi_3-\phi_1)}-\mathcal{C}(1-e^{i(\phi_3-\phi_2)}-e^{i(\phi_2-\phi_1)}+e^{i(\phi_3-\phi_1)})-(1+e^{i(\phi_3-\phi_1)})&=0\\
\end{cases}
\end{aligned}
\end{equation}
Analysing this set of equations, we see that only two phases are independent. Setting $\phi_1=2\pi/3$, $\phi_2=-2\pi/3$ and $\phi_3=0$ leads to a set of fully degenerated equations. \\
$S_{13}(0)=S_{32}(0)=S_{21}(0)=0$ if 
\begin{equation}
2\sqrt{3} \beta -3 \mathcal{C}-1+i \left(4 \alpha -2 \beta +3 \sqrt{3} \mathcal{C}+\sqrt{3}\right) =0.
\label{IsolationEq}
\end{equation}
Solving the previous equation (\ref{IsolationEq}) with respect to the cooperativity $\mathcal{C}$ gives
\begin{equation}
\mathcal{C}=\frac{2 \beta }{\sqrt{3}}-\frac{1}{3} \;\; \text{   and   } \;\;\alpha=-\beta. 
\end{equation}
If $\alpha\neq -\beta$, $\mathcal{C}$ contains an imaginary part leading to complex coupling strengths but we assumed them to be real.  Also, $\mathcal{C}$ has to be positive (such that $g_{ij}\in\mathbb{R}$) and non-zero (else $g_{ij}=0$). The lower bound for $\beta$ is given by
\begin{equation}
\beta>\tfrac{1}{2\sqrt{3}}.
\end{equation}

Let us express the transmission in the counter clockwise direction ($|S_{13}|^2$, $|S_{32}|^2$ and $|S_{21}|^2$)  on resonance as a function of the cooperativity $\mathcal{C}$
\begin{equation}
\begin{aligned}
|S_{13}|^2&=\frac{\kappa_{\text{ex},1}\kappa_\text{ex,3}}{\kappa_1\kappa_3} \frac{1}{(1+\frac{1}{3\mathcal{C}})^2},\\
|S_{32}|^2&=\frac{\kappa_\text{ex,3}\kappa_\text{ex,2}}{\kappa_3\kappa_2}\frac{1}{(1+\frac{1}{3\mathcal{C}})^2},\\
|S_{21}|^2&=\frac{\kappa_\text{ex,2}\kappa_\text{ex,1}}{\kappa_2\kappa_1}\frac{1}{(1+\frac{1}{3\mathcal{C}})^2}.
\end{aligned}
\label{transmissionCirculator}
\end{equation}
We find that, in the case of overcoupled cavities $\kappa_i\approx\kappa_{\text{ex},i}$, the transmission approaches unity with increasing cooperativity.

The symmetrised output noise spectra is computed as in \ref{SI:sec:theory2}. In the limit of overcoupled cavities, $\kappa_i\approx\kappa_{\text{ex},i}$, the noise emitted on resonance at each port is given by
\begin{equation}
N=\frac{1}{2}+\frac{3\mathcal{C}}{(3\mathcal{C}+1)^2} (\bar{n}_\text{m,1}+\bar{n}_\text{m,2})
\end{equation}
with $0<\mathcal{C}<\min\left(\kappa_i/\Gamma_j\right)$. In the limit of large cooperativity, the noise contribution from the mechanical oscillators is entirely suppressed, leaving only vacuum noise of half a quantum.

\end{document}